\documentstyle[12pt]{article} 

\def\r3{{\cal R}_3}

\begin{document}                                                              
\begin{titlepage}
October, 1993 \hfill                 BNL-GGP-1
\begin{center} 
{\LARGE\bf How to make large domains of disoriented chiral condensate}\\
\vspace{1.5cm} 
\large{
Sean Gavin, Andreas Gocksch and Robert D. Pisarski\\
Physics Department, Brookhaven National Laboratory\\
P.O. Box 5000\\
Upton, NY 11973, USA \\
}
\vspace{1.9cm} 
\end{center}
\abstract{Rajagopal and Wilczek have proposed that 
relativistic nuclear collisions can
generate domains in which the chiral condensate is disoriented.  If
sufficiently large ({\it i.e.} nucleus sized), such domains can yield
measurable fluctuations in the number of neutral and charged pions.
However, by numerical simulation of the 
zero-temperature two-flavor linear sigma
model, we find that domains are essentially {\it pion} sized.
Nevertheless, we show that large domains can occur if the
effective mesons masses are much lighter.}
\vfill
\end{titlepage}

In high multiplicity hadronic collisions, it is natural to ask if some
of the observed pions are produced by the coherent decay of a (semi-)
classical pion field \cite{einkova,anselm}.  Bjorken, Kowalski
and Taylor have speculated that events can occur in
which the pion field is oriented along a single direction in isospin
space throughout a large fraction of the collision volume
\cite{bkt,kt}.  They refer to such domains as ``Disoriented Chiral
Condensates'' (DCC) --- disoriented because the pion field vanishes in
the true vacuum.  DCC formation can produce a spectacular event
structure with some regions of the detector dominated by charged
pions and others by neutral pions.  This behavior may have been
observed in Centauro events in cosmic ray collisions \cite{cen}.

Rajagopal and Wilczek (R\&W) \cite{raw} have proposed that the
nonequilibrium dynamics of the chiral transition in relativistic heavy
ion collisions can yield large DCC domains.  They use a linear sigma
model to describe the collective chiral behavior, and make the drastic
assumption that the expansion of the collision fragments ``quenches''
the field.  The initial state is taken to be chirally symmetric, as
appropriate at high temperature, but its time evolution follows the
classical equations of motions at {\it zero} temperature.  R\&W argue
that this quench can amplify the long wavelength pion modes leading to
large DCC domains.

In this Letter we extend the analysis of R\&W to estimate the domain
size and the experimentally relevant fraction of neutral pions
$\r3\equiv n_{\pi^0}/(n_{\pi^+}+n_{\pi^0}+n_{\pi^-})$.  We take
all of the assumptions of R\&W for granted; an analysis incorporating
a more realistic description of heavy ion dynamics will be presented
elsewhere \cite{ggp}.  While our results, like those of R\&W, are the
product of numerical simulations, much of our understanding follows
from the analysis of Boyanovsky {\it et al.} \cite{boy}, who study
domain growth in the context of inflationary cosmology.

To identify a nuclear collision in which pion production is dominated
by domain formation in experiments, one can compare the measured
spectra of neutral and charged pions on an event-by-event basis
\cite{ar}.  The probability for a domain to yield a particular
fraction of neutral pions
$\r3$ is then $P(\r3) = 1/(2\sqrt{\r3})$\cite{com1},
provided that all isospin orientations are equally likely
\cite{bkt,kt,ar,bk}.  In contrast, typical hadronic collisions
produce a binomial distribution of $\r3$ peaked at the isospin
symmetric value of $1/3$ (modulo effects related to the isospin of the
target and projectile).  These distributions are very different,
particularly at small $\r3$.
The size of domains is of central experimental interest.  Nuclear
collisions dominated by a single, large domain can produce a
domain-like $P(\r3)$.  On the other hand, events that produce many
small, randomly-oriented domains result in a binomial distribution.

To describe the evolution of the classical pion field, we use a linear
sigma model that characterizes the chiral order parameter of QCD as an
$O(4)$ vector field $\Phi = (\sigma, \vec \pi)$, where $\sigma$ is an
isosinglet $J^P = 0^+$ field.  The field interacts through a potential
$\lambda({\vec\pi}^2+\sigma^2-v^2)^2/4 - H \sigma$; at zero
temperature the vacuum state is $\langle \Phi \rangle =
(f_{\pi},\vec{0}) $.  The parameters of the model are the 
coupling $\lambda$,
an external field, $H$, which breaks the $O(4)$ symmetry
and the vacuum expectation value of the sigma field in the absence of 
of the external field, $v$.
Our primary interest is in the physically relevant case of strong
coupling, as studied by R\&W.  One takes $\lambda = 20$, $v = 87.4
$~MeV, and $H = (119\, {\rm MeV})^3$, so that the pion decay constant
is $f_{\pi} = 92.5$~MeV and the meson masses are $m_{\pi} = 135$~MeV
and $m_{\sigma} = 600$~MeV.  To establish the role of the $m_{\sigma}$
in domain formation, we compare these results to an extremely weakly
coupled case of $\lambda = 10^{-4}$, $v = 87.4$~MeV, and $H =
(2.03\, {\rm MeV})^3$.  The pion and sigma fields are very light in
this regime, $m_{\pi} = 0.3$~MeV and $m_{\sigma} = 1.8$~MeV, but
$f_{\pi}$ retains the measured value.

Approximating the evolution of the system by a
quench, one assumes that the field is
initially chirally symmetric with $\langle \Phi\rangle = 0$, and
follows its time evolution using the zero-temperature classical
equations of motion.  To simulate the effect of fluctuations
in the high temperature initial state, we distribute the fields
as gaussian random variables with $\langle \Phi 
\rangle = 0$, $\langle \Phi^2 \rangle = v^2/4$ and $\langle \dot \Phi^2
\rangle =v^2/1\,{\rm fm}^2$ following R\&W\cite{raw}.
Pion domains can form in a quench because the chirally symmetric
initial state, $\langle \Phi \rangle = (0,\vec{0})$, is unstable
against small fluctuations in the $T=0$ potential \cite{raw,boy}.  In
essence, the system `rolls down' from the unstable local maximum of
$V(\Phi)$ towards the nearly stable values $\Phi^2 = {f_\pi}^2$.  This
process is known in condensed matter physics as spinoidal
decomposition \cite{boy}.  Long-lived DCC field configurations with
$\vec \pi \neq 0$ can develop during the roll-down period.
The field will eventually settle into stable
oscillations about the unique $H\neq0$ vacuum, $\langle \Phi \rangle =
(f_\pi,\vec{0})$.  Oscillations will then continue until interactions
eventually damp the motion. In the heavy-ion system
a domain can radiate pions preferentially according to its isospin
content.

To be more concrete, we write the equations of motion for the Fourier
components of the pion field in the Hartree approximation 
\cite{raw,boy}:
\begin{equation}
{d^2 \over dt^2}\vec {\pi_{\vec {k}}} \; = \; 
\left( \lambda (v^2 - \langle\Phi^2\rangle) - k^2 \right) 
\vec {\pi_{\vec {k}}}\; .
\end{equation}
\label{eq:motion}
Field configurations with $\langle\Phi\rangle=0$,
$\langle\Phi^2\rangle< v^2$ and momentum $k < \sqrt{\lambda}v $ are
unstable in that they grow exponentially; modes with higher momentum
do not grow.  The $k=0$ mode grows the fastest, with a time scale 
\begin{equation}
\tau_{\rm sp} = \{\lambda (v^2 - \langle\Phi^2\rangle)\}^{-1/2} 
\sim \sqrt{2}/m_\sigma,
\end{equation} 
\label{eq:time}
of order 0.5 fm for $\lambda = 20$. 
The exponential growth of the unstable modes continues
until $\langle\Phi^2\rangle$ reaches $v^2$, when it begins to
oscillate about the stable vacuum.
Rajagopal and Wilczek found that the power $\propto
{\pi^a}_{ \vec {k}} {\pi^a}_{ - \vec{k}}$ in the low
momentum pion modes indeed grows when the exact classical equations of
motion are integrated for $\lambda = 20$.  What
is not clear, however, is whether this observed amplification of power 
is sufficient to create large domains.  

To get a qualitative idea of the domain size, we follow Boyanovsky
{\it et al.} \cite{boy} and estimate the contribution of the unstable
modes to the pion correlation function.  For small times, we treat
$\langle \Phi^2 \rangle=v^2/4$ as a constant, so that
$\vec {\pi}_{\vec{k}}(t)=\vec {A}_{\vec{k}}e^{t/\tau_k}+
\vec {B}_{\vec{k}}e^{-t/\tau_k}$,
with $\tau_k^{-1}=\sqrt{3 \lambda v^2 /4+ k^2}$.  
The averages $\langle \langle A_k A_{k^{\prime}}\rangle \rangle,\langle
\langle A_k B_{k^{\prime}}\rangle \rangle$
over the initial conditions
can be computed explicitly\cite{ggp}.
For $t\gg\tau_{\rm sp}$, we find
\begin{equation}
\langle \langle \vec{\pi(x,t)}\vec{\pi(0,t)}\rangle \rangle
\; \approx \; \langle \langle\vec{\pi(0,t)}^2\rangle \rangle \; {\rm exp} 
\left\{- x^2 /8\tau_{\rm sp}t \right\}\; .
\end{equation}
suggesting that domains grow roughly as
$\xi_D \sim \sqrt{8 \tau_{\rm sp} t}$. 
We therefore expect typical domains to reach a size $\sim 
\sqrt{8}\tau_{\rm sp}$, since they can only grow until the time $t \sim
\tau_{\rm sp}$ when the field reaches the minimum of the potential.
If the coupling were weak and the sigma meson light, it would take a
long time to roll down to the bottom of the potential, so that domains
would have plenty of time to grow.  However, for the more realistic
strongly-coupled case, the rolldown is very rapid and the domains are
small, perhaps $\sim 1.4$~fm, which is comparable to the Compton
wavelength of the pion.

We confirm this intuition through numerical simulations.
Specifically, we simulate the model on a three dimensional $10^2
\times 40$ lattice with a spacing $a = 1$~fm, computing its evolution
using a simple leap-frog algorithm.  To extract domain-size
information more easily, we employed an asymmetric lattice and studied
the pion field averaged over the short dimensions,
$\vec{\pi}_L(z,t) = \sum_{x,y}\vec{\pi}(x,y,z,t)/(10)^2$.
Our simulations qualitatively reproduce the time evolution of the
momentum space quantity $\pi^a_{\vec k}\pi^a_{-\vec{k}}$
reported by R\&W for a symmetric lattice at $\lambda = 20$.  In
configuration space, $\pi_L$ starts out zero with small gaussian
fluctuations.  Fig. 1 shows the profile of $\vec{\pi_L}$ at a time $t =
100$ fm in weak coupling, $\lambda = 10^{-4}$, and the same
profile at a time $t = 30$ fm in strong coupling, $\lambda = 20$.  For
weak coupling, regions in which the field is slowly varying about some
nonzero value --- domains --- are evident.  On the other hand, in
strong coupling the pion field is oscillating with small amplitude
about zero.  We have also measured the pion correlation function: it
is long ranged in weak coupling, but drops to zero within a few
lattice spacings in strong coupling \cite{ggp}.  We also varied the
time at which correlations were measured, and were never able to find
large domains in strong coupling.

We also studied the distribution of the fraction of neutral pions
$\r3=\langle(\pi^3)^2\rangle/ \sum\langle(\pi^a)^2\rangle$.
Histograms of $\r3$ were obtained by evolving $500$ independent
configurations in time to $t = 150$ and $t=30$ fm at weak and strong
coupling respectively. These are shown in
Figure 2.
In weak coupling the distribution is far from binomial and favors
small $\r3$ like
the DCC distribution $P(\r3) = 1/(2\sqrt{\r3})$. In strong coupling,
however, the distribution is clearly binomial, peaked about the
expected value of $1/3$.

In summary, we find that distributions of the fraction of neutral
pions do not reflect domain formation for the two-flavor linear sigma
model assuming the R\&W quench.  The $\r3$ distribution
can reflect domain formation only when the domain size $\xi_D$ exceeds
the system size $R$.  We find that DCC-like behavior dissappers for
couplings exceeding $\lambda\sim 10^{-2}$ in our $10^2 \times 40$
lattice, in accord with (3) \cite{ggp}.  However, a system the size of
a gold nucleus can exhibit domain-like behavior for $\lambda <
8/(vR)^{2}\sim 1$.

A necessary condition for large domains of DCC's to arise is 
that there is a
light particle about --- the dynamics of a quench does not give us a
large distance scale for free.  However, heavy ion collisions can
create a high temperature state that cools over time scales $\gg
\tau_{\rm sp} \sim 0.5$~fm. The quench might therefore not be a
a very good approximation.  {\it If} the equilibrium thermodynamics of the
QCD phase transition is near a critical point, then near the
transition temperature there is automatically a light field about,
providing a natural mechanism for the growth of large domains.  In
forthcoming papers we will show how  such a critical point is present
in the QCD phase diagram, and analyze the phenomenology of the DCC
domains produced thereby \cite{ggp}.


We thank K. Rajagopal for discussions, and for sending us a copy
of his thesis.
This work was supported by a DOE grant at 
Brookhaven National Laboratory (DE-AC02-76CH00016).

\pagebreak

\section*{Figure Caption}

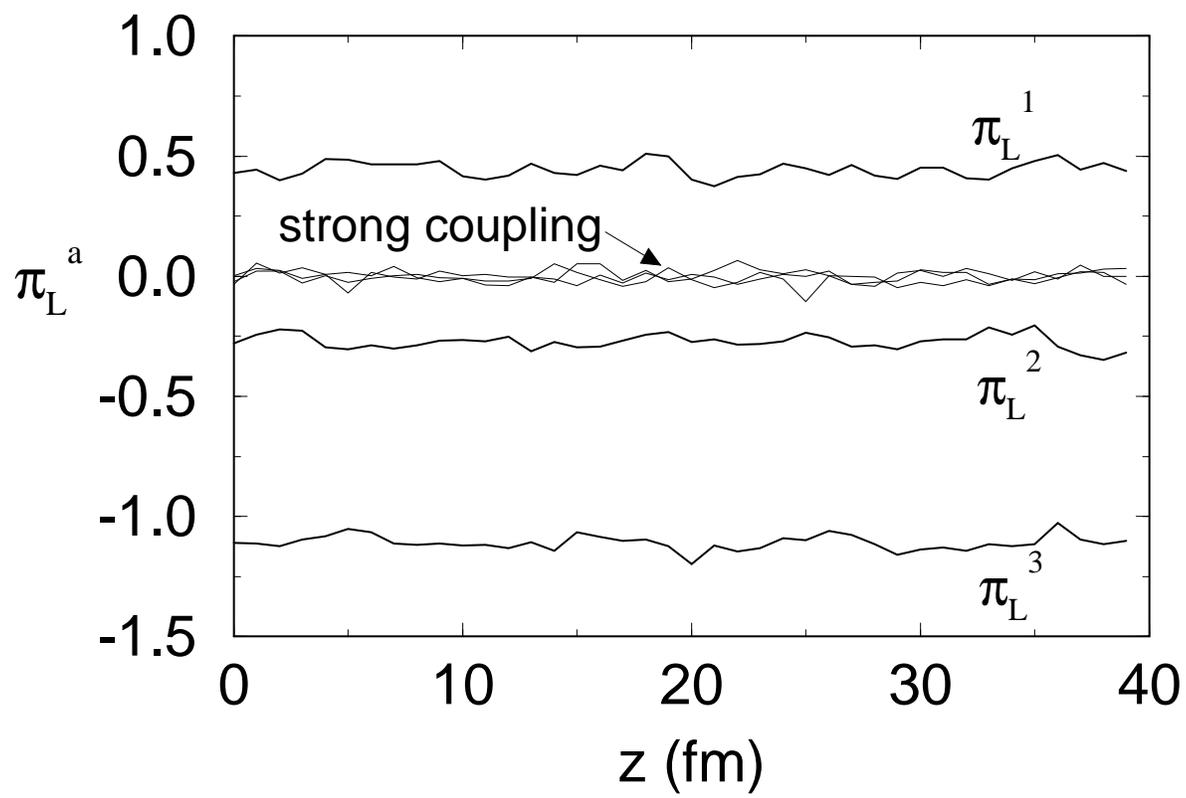
\noindent {\bf Figure 1}:
Snapshot of the pion field, $\vec{\pi}_L$ on a $10^2 \times 40$ lattice 
taken in weak coupling ($\lambda = 10^{-4}$) at $t = 100$ fm and at
$t=30$ fm at strong ($\lambda = 20$) coupling. Note how at strong 
coupling the three components of $\vec{\pi}_L$ fluctuate around zero.
The labeled curves show $\vec{\pi}_L$ at weak coupling. All three
components are non-zero and essentially constant throughout the 
length of the box.

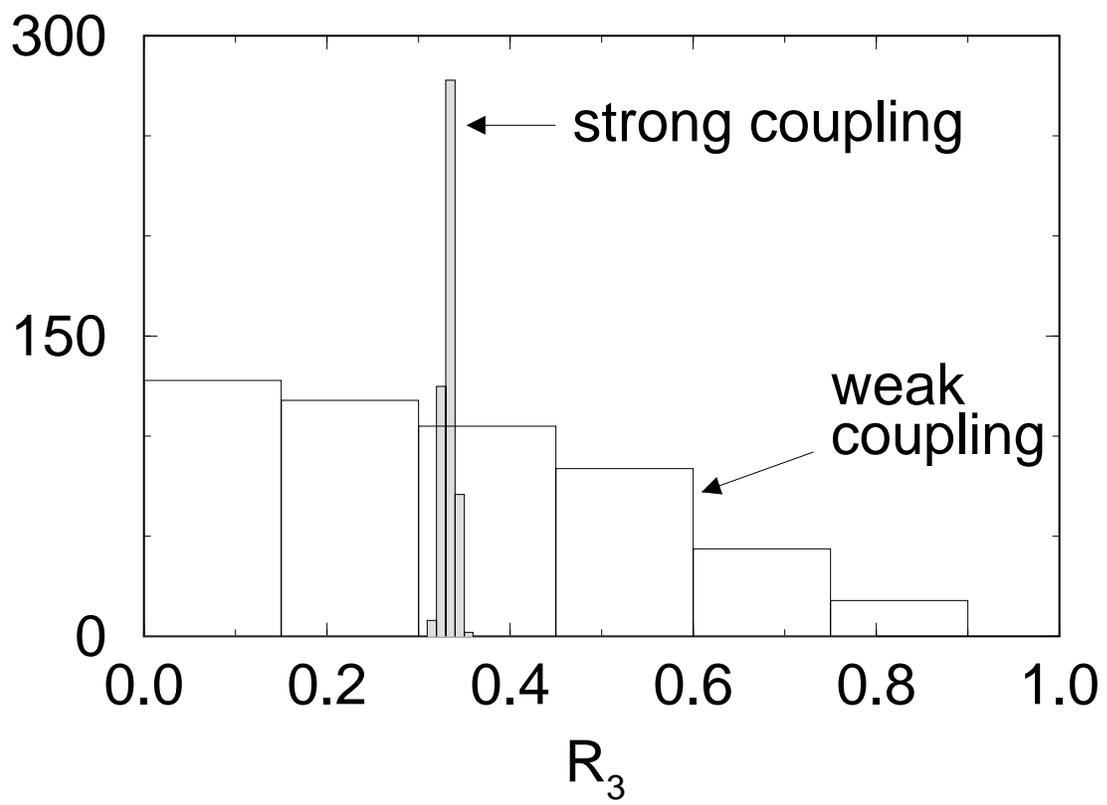
\noindent {\bf Figure 2}:
The distribution of the fraction of neutral pions, $\r3$
on a $10^2 \times 40$ lattice 
in weak coupling ($\lambda = 10^{-4}$) at $t = 100$ fm and at
$t=30$ fm at strong ($\lambda = 20$) coupling. In each case $500$
different initial configurations were used to generate the histograms.
At strong coupling the distribution is narrowly peaked around 
$\r3 =1/3$ wheras at weak coupling it is broad and clearly giving
more weight to small values of $\r3$.
\pagebreak


\begin{thebibliography}{99}
\bibitem{com1}
We have chosen the $3$ direction in isospin to be that for neutral pions,
but there is nothing special in this choice.  For
example, if ${\cal R}_1$ is the fraction of pions in the isospin
$1$ direction, then 
the distribution in ${\cal R}_1$ is $1/(2 \sqrt{{\cal R}_1})$.  
This is obvious, since by an isospin rotation we can relabel the $3$ direction
as the $1$ direction.
\end{thebibliography}
\end{document}